\documentclass[epj]{svjour}
\usepackage{graphics}
\begin{document}
\title{Magneto-elastic Coupling in the Layered Manganite
La$_{1.2}$Sr$_{1.8}$Mn$_2$O$_7$}
\author{C.N. Brosseau\inst{1}, M. Poirier\inst{1}, R. Suryanarayanan\inst{2}, G. Dhalenne\inst{2}
\and A. Revcolevschi\inst{2}}
\offprints{}          % Insert a name or remove this line
\institute{Centre de Recherche sur les Propri\'et\'es
\'Electroniques de Mat\'eriaux Avanc\'es and D\'epartement de
Physique, Universit\'e de Sherbrooke, Sherbrooke, Qu\'ebec, Canada
J1K 2R1. \and Laboratoire de Physico-Chimie de l'Etat Solide, UMR
8648, B\^at. 414, Universit\'e Paris-Sud, 91405 Orsay C\'edex,
France}
\date{Received: date / Revised version: date}
\abstract{We have studied the magneto-elastic coupling in the
double layered Mn perovskite La$_{1.2}$Sr$_{1.8}$Mn$_2$O$_7$ with
an ultrasonic velocity technique. The temperature profile of both
the in-plane and out-of-plane longitudinal velocities showed a
large stiffening anomaly below the insulating paramagnetic to
metallic ferromagnetic transition. Magnetic fluctuations effects
consistent with the layered structure are evidenced as a
frequency dependent velocity softening above the transition. The
magneto-elastic coupling has been studied in magnetic field
values up to 8 Tesla: the observations are consistent with a
substantial magnetic anisotropy and a ferromagnetic order
parameter with moments lying in the layers.
 % end of PACS codes
} %end of abstract
\titlerunning{Magneto-elastic Coupling in the Layered Manganite...}
\authorrunning{Brosseau et al.}
\maketitle
\section{Introduction}
\label{intro} Doped perovskite manganites La$_{1-x}$A$_x$MnO$_3$
(A = Sr, Ca) are continuing to attract much attention mainly
because they are showing a wide variety of magnetic-field induced
phenomena like the colossal magnetoresistance (CMR). In these
materials, although the double- exchange mechanism \cite{Zener}
serves as a starting point to explain the semiconductor
(paramagnetic) to metallic (ferromagnetic) phase transition and
the associated CMR, the Jahn-Teller effect and strong
electron-phonon interaction must be included into the model
\cite{Millis}. Recent research in these mixed valence manganites
has expanded to include the search for materials with other
structure types exhibiting CMR.  For example, one can adjust the
dimensionality of the La$_{1-x}$Sr$_x$MnO$_3$ perovskites by
inserting a blocking layer (La$_{1- x}$Sr$_x$)$_2$O$_2$ between
MnO$_2$ sheets to yield the Ruddlesden Popper series of compounds
with reduced dimension $n$, (La$_{1-
x}$Sr$_x$)$_{n+1}$Mn$_n$O$_{3n+1}$
\cite{Mahesh,Battle,Laffez,Asano}. The compounds
La$_{1-x}$Sr$_x$MnO$_3$ belong to this family with $n = \infty$.
The bi-layer perovskite (La$_{1-x}$Sr$_x$)$_{n+1}$Mn$_2$O$_7$ in
which two MnO$_6$ layers are stacked with
(La$_{1-x}$Sr$_x$)$_2$O$_2$ layers along the $c$ axis of the
structure is obtained for $n = 2$.  The reduced dimensionality
has been shown to have interesting consequences on the physical
properties of these compounds. The material is antiferromagnetic
(AFM) for $x = 0$, while it is ferromagnetic (FM) and exhibits a
metal insulator (MI) transition at T$_c$ in the region $0.2 \leq
x \leq 0.4$. The maximum T$_c$ = 125 K is obtained for $x$ $\sim$
0.4 and a large value of CMR is obtained at temperatures near and
far away from T$_c$ \cite{Morimoto,Prellier}.  One of the most
distinct features for the bilayered manganites is its anisotropic
characteristics in charge transport and in magnetic properties.
They offer then a rich opportunity to explore the interplay
between spin, charge and lattice degrees of freedom in reduced
dimensions. In these materials the magnetization process causes
not only a large negative magnetoresistance but also a gigantic
anisotropic lattice striction that reveals a complex
unconventional spin-charge- orbital coupling
\cite{Kimura,Ogasawara,Argyriou1}. This magneto-elastic coupling
is dependent on the magnetic structure and critical phenomena and
its role in the mechanism of CMR enhancement has to be clarified.

Neutron diffraction studies \cite{Hirota} have revealed that the
low temperature magnetic phase of these compounds consists of
planar ferromagnetic (FM) and A-type antiferromagnetic (AFM)
components, indicating a canted AFM ordering. For $x$ = 0.4 the
canting angle betwen planes is 6.3$^\circ$ (nearly planar FM).
Hirota {\it et al.} \cite{Hirota} also found that the A-type AFM
ordering remained above T$_c$ and that it showed an anomalous
exponential decrease to T$_N$ $\sim$ 200 K.  These authors
suggested that this AFM phase could play a significant role in the
enhancement of CMR effects in this layered Mn system.  In another
neutron experiment \cite{Argyriou}, it was shown that the
ferromagnetic transition at T$_c$ which is exhibiting critical
scattering and a divergent coherence length is accompanied by 2D
ferromagnetic correlations over a wide temperature range.  These
FM correlations were found to fluctuate rapidly close to T$_c$.
It was then concluded that the layered manganites provide a
unique opportunity to examine in detail the mechanism of the
crossover between 2D and 1D magnetism.

In this paper, we investigate with an ultrasonic velocity
technique the magneto-elastic coupling in the layered compound
La$_{1.2}$Sr$_{1.8}$Mn$_2$O$_7$ which is showing the maximum
T$_c$. Because of a strong coupling between spins and
longitudinal acoustic waves, large anomalies are observed on the
temperature profile of the acoustic velocity. These anomalies
have been studied for different crystal directions in magnetic
field values in the range 0-8 Tesla. These data will be discussed
in relation with the magnetic structure and with the existence of
2D fluctuations on a wide temperature range. A
temperature-magnetic field diagram is proposed.

\section{Samples and Experiment}
\label{sec:1} Single crystals of La$_{1.2}$Sr$_{1.8}$Mn$_2$O$_7$
were grown from the melt by a conventional floating-zone method
using a mirror furnace \cite{Revco}. The compounds crystallize in
the body-centered tetragonal structure $I4/mmm$ ($D^{17}_{4h}$)
\cite{Mitchell}. The crystals could easily be cleaved to lead to
shiny surfaces in the $ab$ plane with the $c$ direction
perpendicular to it. Neutron diffraction experiments on cm size
crystals confirmed the single crystallinity of the sample. For the
present experiment, the quality of the single crystal, sliced
from a bigger sample, was also checked with optical polarizing
microscopy. EDX analysis indicated a homogeneous sample without
any extra phases (limit of detection 1$\%$) and further the
composition was found to be close to that of the starting
material. The ultrasonic velocity is measured with a pulsed
acoustic interferometer \cite{Poirier} yielding a sensitivity in
relative velocity variation better than 1 ppm. Although the
crystals have large dimensions, so large parallel faces that are
well suited for pulsed echo ultrasonic measurements, they cannot
be used directly. Indeed, the magneto-elastic coupling is so
strong and the attenuation (decibel/mm) in the MHz range so high
that we had to use much smaller crystals and, doing so,we had to
insert a delay line (CaF$_{2}$) to separate the ultrasonic
echoes. The crystal used in our experiment had dimensions 0.95,
1.15 and 0.75 mm respectively along the $a$, $b$ and $c$ axes.
One parallel face of the crystal is first glued on the surface of
a CaF$_2$ delay line (buffer length $\sim$ 7 mm). Then, a
LiNbO$_3$ piezoelectric transducer bonded on the other parallel
face generates longitudinal waves, at 30 MHz and odd overtones,
that will propagate through the crystal-delay line ensemble and
will be detected by a second piezoelectric transducer at the
other end of the buffer. The longitudinal waves can be propagated
either along the $a$ (or $b$) axis or the perpendicular direction
$c$. Transverse waves have not been used for this study since they
could not be isolated easily from other modes in the ultrasonic
experiment because of the small dimensions of the crystal.

The temperature profile of the velocity is obtained by monitoring
the phase ($\phi$) of the transmitted signal as a function of the
temperature. However, since two terms,  $\phi = k_1 l_1 + k_2
l_2$ where $k_1$, $k_2$, $l_1$ and $l_2$ are respectively the
ultrasonic wave vectors and the lengths of the crystal (1) and
the delay line (2), contribute to the total phase of the
transmitted signal, it is necessary to substract the delay line
contribution ($k_2 l_2$) to isolate the crystal one ($k_1 l_1$).
The former is measured in a pulsed reflection experiment in the
delay line only at the same frequency and for identical
experimental conditions.  Our ultrasonic data have not been
corrected to take into account thermal expansion (striction
\cite{Kimura,Argyriou1}) effects ($\Delta l_1 \over l_1$) since
these are generally orders of magnitude too small compared to the
measured relative velocity variation ($\Delta v \over v$). The
absolute values of the longitudinal velocities are around
v$_{a,b}$ = 4300 and v$_c$ = 3400 m/sec. No anisotropy was found
between the $a$ and $b$ directions. The temperature profile of
the velocities is obtained in the range 4-200 K in magnetic field
values between 0-8 Tesla. The temperature is monitored either
with a Si diode or a carbon glass sensor and stabilized with a
LakeShore controller. For reasons of excellent reproducibility
during thermal cycling, silicon seal was used for all bonds
between the crystal, the transducers and the buffer; no velocity
data can be obtained above 200 K where the silicon seal presents
a phase transition.

\input epsf.tex \vglue 0.4 cm
\epsfxsize 8 cm
\begin{figure}
\centerline{\epsfbox{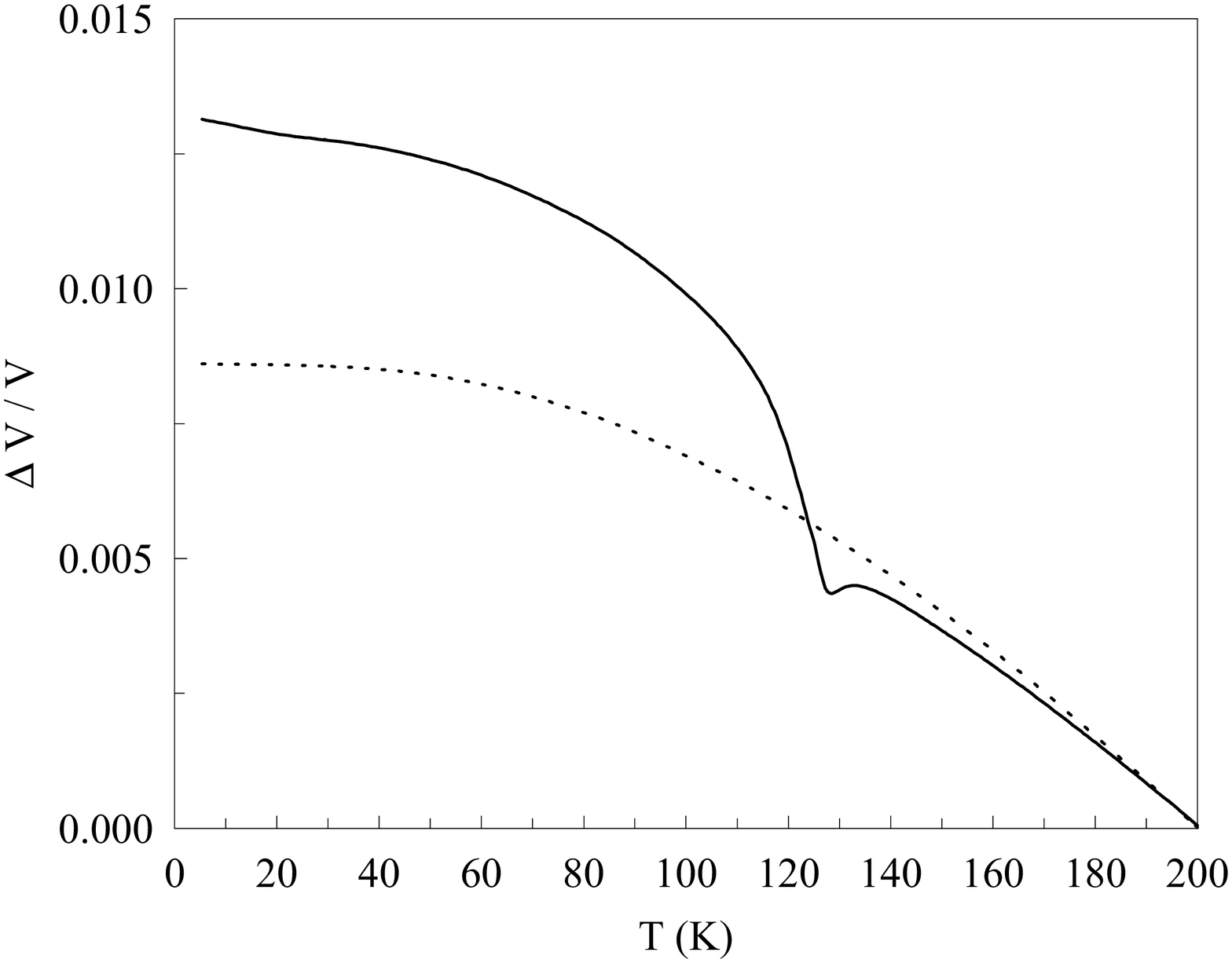}} \caption{Relative variation of
the longitudinal velocity as a function of temperature obtained
at 165 MHz: crystal/buffer (full line) and buffer (dotted line).}
\label{}
\end{figure}
\section{Results and Discussion}
\label{sec:2} In figure 1 we present an example of the
temperature profile of the variation of the velocity (relative to
its value at 200 K) for 165 MHz longitudinal waves propagating
either trough the crystal/buffer ensemble (full line) or the
buffer alone (dotted line). The velocity in the buffer shows the
usual monotonous behavior which is related to the anharmonic
contribution.  This means that all the features observed on the
crystal/buffer profile are due to magneto-elastic effects in the
La$_{1.2}$Sr$_{1.8}$Mn$_2$O$_7$ crystal. When the buffer
contribution ($k_2l_2$) to the overall phase ($\phi$) is
substracted, we get the temperature profile of the crystal
velocity that is shown in figure 2 for 165 MHz longitudinal waves
having their wave vector $k$ parallel to the $a$ (or $b$) axis or
perpendicular to it ($k \parallel c$). When the temperature is
decreased below 200 K for both crystal directions, the velocity
first increases smoothly but, in the vicinity of the phase
transition temperature T$_c$, a small softening of the velocity
is observed before an important stiffening progressively sets in
below T$_c$. At low temperatures the smooth behavior is recovered
although a small slope increase is seen around 20 K. Differences
between the two directions are a larger stiffening anomaly below
T$_c$ and a more pronounced softening around T$_c$ when waves
propagate along the $a$ (or $b$)axis. As it is observed on the
transport properties \cite{Prellier}, when a 8 Tesla magnetic
field is applied along $a$ (or $b$), the stiffening anomaly is
shifted to higher temperatures as shown in Fig.2. Such a large
acoustic anomaly ($\sim 2\%$ in amplitude) is typical of what is
found in strong magnetic materials and we thus suggest to relate
it to a magneto-elastic coupling at the paramagnetic-ferromagnetic
transition.  Indeed, since quasi-particle screening effects are
expected to be orders of magnitude smaller and to rather yield a
softening anomaly when entering the metallic state below T$_c$,
the charge degrees of freedom cannot explain the features
observed here.

\input epsf.tex \vglue 0.4 cm
\epsfxsize 8 cm
\begin{figure}
\centerline{\epsfbox{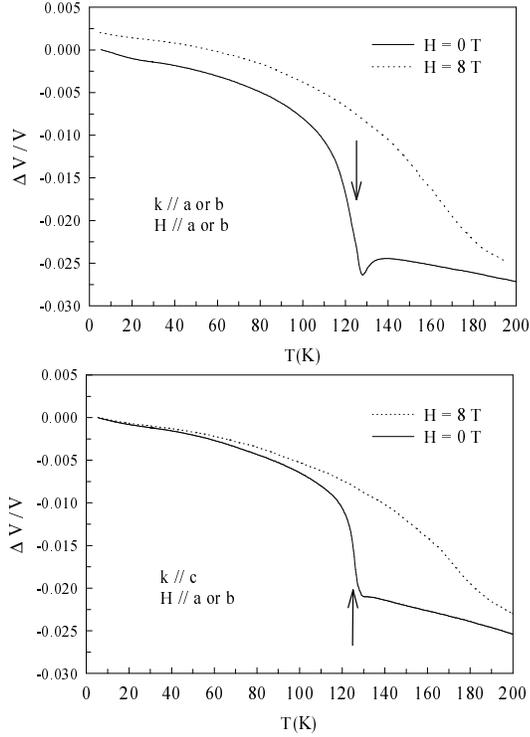}} \caption{Relative variation of
the longitudinal velocity as a function of temperature obtained
at 165 MHz: $k \parallel$ $a$ or $b$ (upper panel) and $k
\parallel c$ (lower panel). The arrows indicate the transition
temperature T$_c$.} \label{}
\end{figure}

In order to interpret the velocity data presented in Fig.2, we
need to define correctly the transition temperature T$_c$. This
can be done most easily on the temperature profile of the
resistivity $\rho (T)$ where the maximum rate ($d\rho /dT$)
generally defines T$_c$. We present in figure 3 the microwave
resistivity along $c$ as a function of temperature obtained at 17
GHz on the same crystal. The semiconductor metal transition is
characterized by an abrupt decrease of the resistivity at T$_c$ =
125 K where a dashed line indicates the maximum variation of the
resistivity. The temperature profile and the transition
temperature are similar to the DC ones reported in the literature
\cite{Prellier}. In the ultrasonic experiment, this value of T$_c$
corresponds to the temperature of maximum stiffening rate
observed below the weak softening peak as indicated by an arrow
in Fig.2.

\input epsf.tex \vglue 0.4 cm
\epsfxsize 8 cm
\begin{figure}
\centerline{\epsfbox{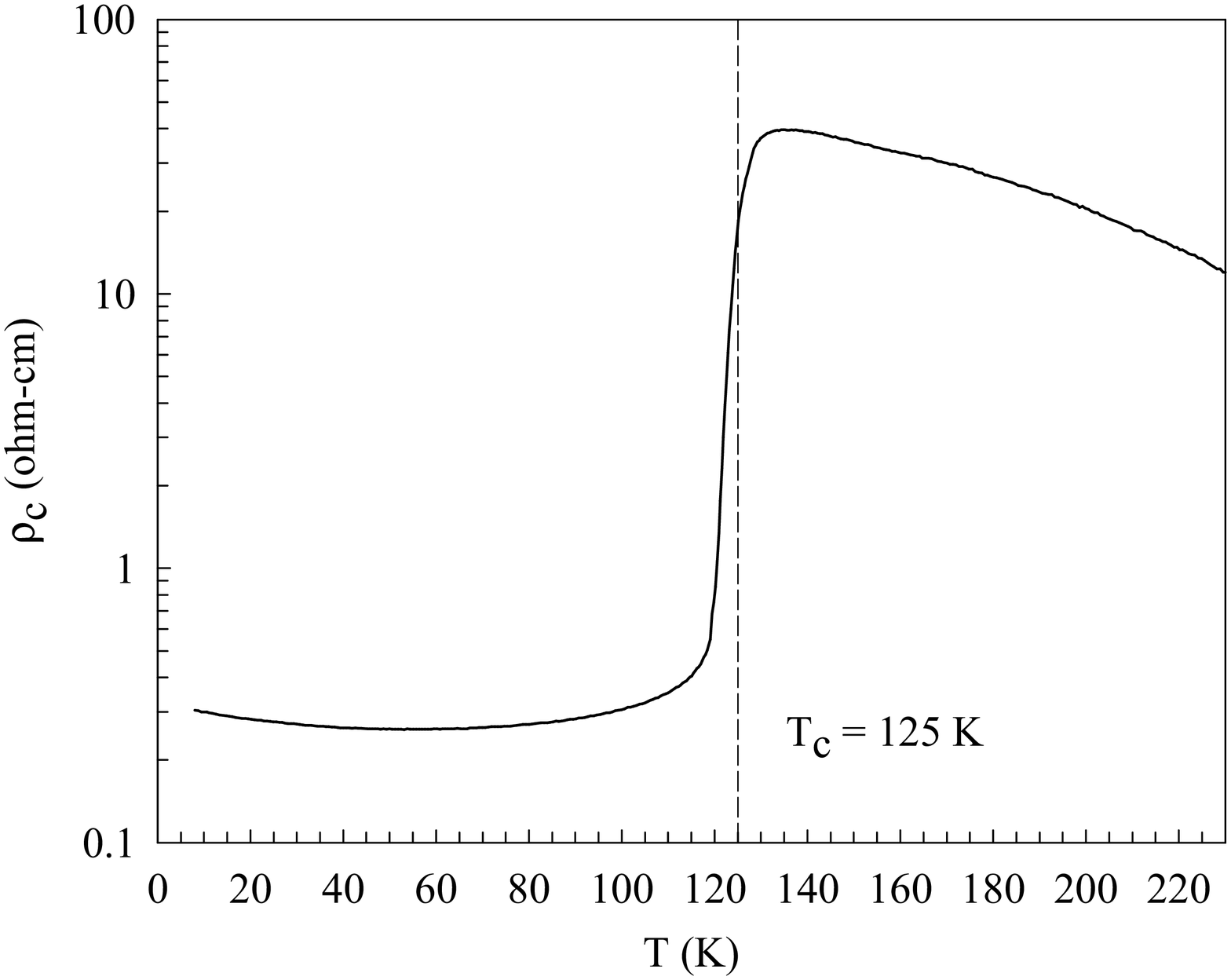}} \caption{Microwave resisitivity
(17 GHz) along $c$ for the layered manganite
La$_{1.2}$Sr$_{1.8}$Mn$_2$O$_7$ as a function of temperature. The
dashed line indicates the transition temperature T$_c$.} \label{}
\end{figure}

For T$<$T$_c$ the stiffening anomaly is likely due to a coupling
between ferromagnetic spins and longitudinal acoustic phonons.
Indeed, the stiffening anomalies in zero field shown in Fig.2 are
coherent with this picture: a spontaneous magnetization
progressively builds up below T$_c$ and this gives an increase of
$\Delta v / v$. We have tried to isolate the magneto-elastic
coupling by substracting the anharmonic contribution which is
found by extrapolating to low temperatures the profile observed
for T$>$T$_c$.  The result is presented in figure 4 for both
crystal directions. A stiffening anomaly which mimics the
magnetization $M$ of the ferromagnetic phase is obtained.  At
T$_c$, the stiffening appears to be sharper along the $c$ axis
than along $a$ (or $b$) and the anomalies have different
amplitudes, being larger along the $ab$ plane. A change of slope
below 20 K is clearly observed on both curves. Its origin could
not be identified as it is hardly modified by a magnetic field
and, thus, it will not be discussed further. The softening
observed above T$_c$ is due to magnetic fluctuations and this
will be discussed shortly.

\input epsf.tex \vglue 0.4 cm
\epsfxsize 8 cm
\begin{figure}
\centerline{\epsfbox{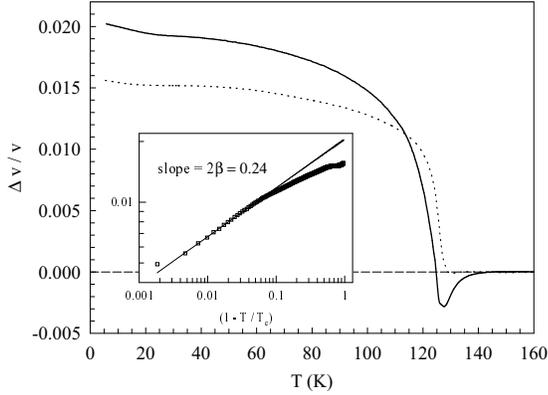}} \caption{Magneto-elastic
contribution to the velocity profile at 165 MHz in zero field: $a$
or $b$ axis (full line) and $c$ axis (dotted line). Inset: $c$
axis data as a funtion of the reduced temperature (1 - $T\over
T_c$) on a log-log scale;  the straight line represents the best
fit in the critical region with $\beta$ = 0.12. } \label{}
\end{figure}

The stiffening anomaly can be explained by a magneto-elastic
coupling according to the following scenario. We consider a
coupling between the strain $e$ and the order parameter $Q$
(spin-spin correlation function) and the energy term is
phenomenologically expanded in powers of these parameters
\cite{Pouget}: a stiffening anomaly below T$_c$ is obtained if we
consider a bi-quadratic coupling energy term $\lambda e^2Q^2$. In
most favorable cases where a single ferromagnetic component is
present in the low temperature phase and where domain effects are
absent, the measurement of the elastic constant $C$, so the
velocity $v$ ($C = \rho v^2$, $\rho$ being the mass density), can
reveal the temperature dependence of the order parameter. The
curves shown in Fig.4 are in agreement with such a scenario.
Since the largest anomaly is observed for in-plane propagation, a
larger coupling constant $\lambda$ is expected. However,
ferromagnetic domains are likely present in
La$_{1.2}$Sr$_{1.8}$Mn$_2$O$_7$ crystals and a AFM component has
also been identified in neutron experiments \cite{Hirota} at low
temperatures; we thus have to be careful to extract the order
parameter from these curves, especially in the critical
temperature region. Indeed, the data presented in figure 2 reveal
that a 8 Tesla magnetic field oriented along $a$ (or $b$)
modifies differently the low temperature amplitude of the anomaly
according as the wave is propagating parallel or perpendicular to
the plane. For the parallel case, a further stiffening likely due
to domain wall effects is observed in the ferromagnetic phase
when no effects are detected for the perpendicular one.

Longitudinal waves can interact most easily with domain walls that
are perpendicular to the elastic polarization. For a tetragonal
structure where the Mn magnetic moments are ordered
ferromagnetically along the $a$ or $b$ axes, four domains are
expected in zero field and the walls should run preferentially
along the $c$ axis. The application of a magnetic field will
affect the domain structure and the velocity should be changed
accordingly. These magnetic field effects investigated at 10 K,
far below the critical region, are presented in figure 5 for
different configurations of wave propagation and field
orientation. In the upper panel we are considering waves
propagating along $a$ (or $b$). When the field is along the same
direction (easy plane), the velocity first softens weakly, reaches
a minimum around 0.12 Tesla, then stiffens and saturates above 0.4
Tesla. The minimum could be the result of an increase softening
due to domain wall motion before stiffening is obtained when the
90 degrees domains rotate to align along the field. When the field
is rather along the hard axis ($c$), no minimum is observed and
saturation is obtained at a much higher field value around 1.3
Tesla. This is consistent with a large magnetocrystalline
anisotropy with the hard axis along the $c$ direction; as the
domain structure is not modified by a perpendicular field, no
softening is observed for this configuration. In the lower panel
of Fig.5, we notice that longitudinal waves propagating along $c$
are not affected if the field is along $a$ (or $b$) since
interaction with domain walls is not efficient. For field along
$c$ however, softening is obtained when the moments are forced to
leave the easy plane with a maximum obtained around 1.4 Tesla and
saturation is observed above 2 Tesla when all the moments are
aligned in the field. The negative value obtained at saturation
is likely related to the demagnetization field. This dependence
of the velocity in the ferromagnetic phase is fully coherent with
the magnetic structure and the field behavior observed by others
\cite{Sharma} although the absolute values are somewhat
different. In these measurements however, the possible role
played by a small AF component is difficult to evaluate.

\input epsf.tex \vglue 0.4 cm
\epsfxsize 8 cm
\begin{figure}
\centerline{\epsfbox{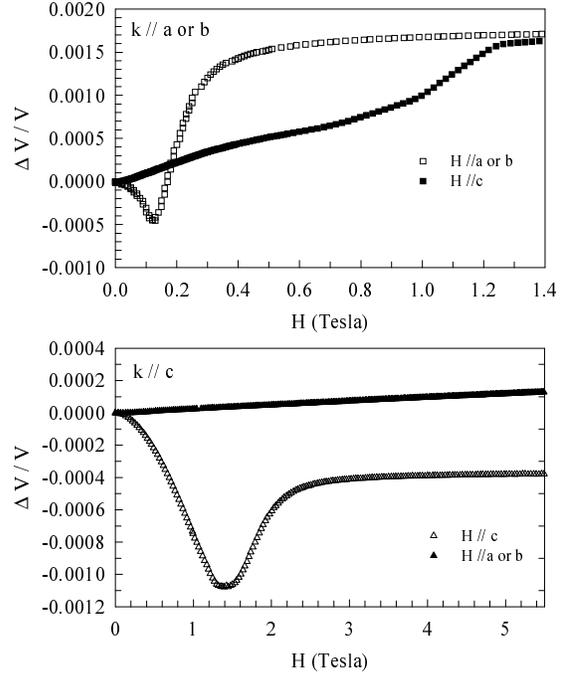}} \caption{Variation of the
velocity relative to its value at zero field at 165 MHz as a
function of magnetic field at 10 K. Upper panel, wave vector $k
\parallel a$ or $b$; lower panel, $k \parallel c$. } \label{}
\end{figure}

The effect of domains on the velocity data complicates the
determination of the temperature profile of the order parameter,
especially the critical behavior near the phase transition
temperature. If a bi-quadratic magneto-elastic energy term is
used, one can show that the relative velocity change is
proportional to the square of the order parameter \cite{Pouget}.
Now, if we consider the $c$ axis velocity data of Fig.4 as being
free of magnetic domains effects, the critical region near the
phase transition can be fitted reasonably well to the following
relation, $({\Delta v \over v})^{1 \over 2}\sim Q(T) \sim
(T_{c}-T)^{\beta}$, with parameters T$_c$ = 125.5 $\pm$ 0.2 K and
$\beta$ = 0.12 $\pm$ 0.01 (see inset of Fig.4). The small value of
the exponent is consistent with the presence of strongly two
dimensional fluctuations below T$_c$ as predicted by the 2D Ising
model ($\beta$ = 0.125). If this weak value of $\beta$ agrees
with the neutron scattering data of Osborn et al. \cite{Osborn}
and Rosenkranz et al. \cite{Rosenkranz}, it contradicts the 3D
Heisenberg value (0.35) of Chatterji et al. \cite{Chatterji}.
However, our results indicate clearly that 2D magnetic
fluctuations are dominating the critical temperature region. At
much lower temperatures, the $c$ axis data of Fig.2 (lower panel)
reveal that, although a magnetic field shifts the phase
transition to higher temperatures, it has no effect on the low
temperature value of the order parameter $Q(0)$. Such a field
independent order parameter $Q(0)$ has also been observed
previously in low-dimensional AFM systems \cite{Allen,Poirier1}.

\input epsf.tex \vglue 0.4 cm
\epsfxsize 8 cm
\begin{figure}
\centerline{\epsfbox{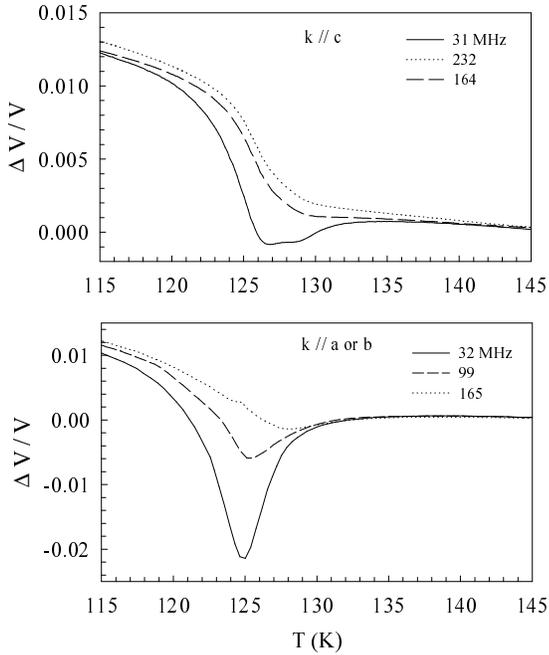}} \caption{Relative variation of the
velocity as a function of temperature near T$_c$ at different
frequencies in the MHz range and in zero field. Upper panel, wave
vector $k \parallel c$; lower panel, $k \parallel a$ or $b$.}
\label{}
\end{figure}

The presence of 2D magnetic fluctuations can also be inferred
directly from the velocity data presented in figure 4. Indeed, the
velocity softening observed just above T$_c$ for in-plane
propagation can be attributed to fluctuations and the fact that
this softening is hardly seen for $c$ axis propagation implies
that the fluctuations possess a strong 2D character. In fact, the
fluctuation regime for T $>$T$_c$ is highly dependent on the
frequency in the MHz range. This can be noticed in figure 6 where
we present the velocity softening obtained at different
ultrasonic frequencies near the transition. For in-plane
propagation (lower panel), the amplitude of the softening
increases dramatically when the frequency is decreased from 165
to 32 MHz (it reaches more than 2$\%$ at the lowest frequency,
the same value as the stiffening due to the order parameter below
T$_c$) and its maximum comes very close to T$_c$ (125 K). For
out-of-plane propagation (upper panel), a similar effect is
obtained but the softening amplitude is much smaller by at least
one order of magnitude. This clearly indicates that, for T
$>$T$_c$, 2D fluctuations are strongly coupled to acoustic phonons
and that their relaxation time is larger than $10^{-7}$ sec. The
range of temperatures over which these fluctuations are extending
is difficult to determine since we do not know exactly the
temperature profile of the velocity without the effects of spins.
According to the data of Fig.6 these fluctuations can be detected
at least up to 150 K. This frequency dependence is consistent
with the ferromagnetic character of the 2D fluctuations since
these are expected to be enhanced as the wave vector $q
\rightarrow 0$.

\input epsf.tex \vglue 0.4 cm
\epsfxsize 8 cm
\begin{figure}
\centerline{\epsfbox{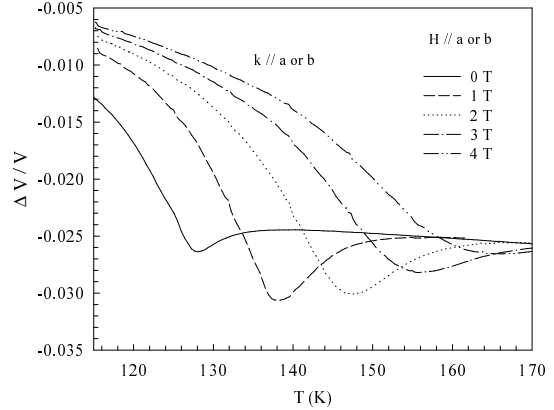}} \caption{Relative variation of the
velocity at 165 MHz as a function of temperature near T$_c$ for
different magnetic field values (H $\parallel a$ or $b$): wave
vector $k$ oriented along $a$ or $b$ axes. } \label{}
\end{figure}

It is also well known that the application of a magnetic field
can accentuate the fluctuations in low-dimensional magnetic
systems. This can be seen in figure 7 where we indicate how the
critical region for in-plane propagation is modified by a
magnetic field oriented in the plane. In fact, the field is not
only shifting the transition temperature to higher values, but it
also increases its width. Besides, the amplitude of the softening
increases first with field up to 1 Tesla and then decreases
smoothly as the critical region becomes wider and wider. The same
trend is also observed for out-of-plane propagation with much
smaller effects. This behavior is again consistent with the
presence of 2D ferromagnetic fluctuations strongly coupled to
acoustic phonons over a wide temperature range above T$_c$.
Whether the character of the fluctuations is 2D Ising is not fully
established since an effective finite-size 2D XY model has been
suggested to explain quantitatively the critical properties above
and below T$_c$ \cite{Rosenkranz}, although the exponent
predicted by this model seems too large (0.23) \cite{Bramwell}.

\input epsf.tex \vglue 0.4 cm
\epsfxsize 8 cm
\begin{figure}
\centerline{\epsfbox{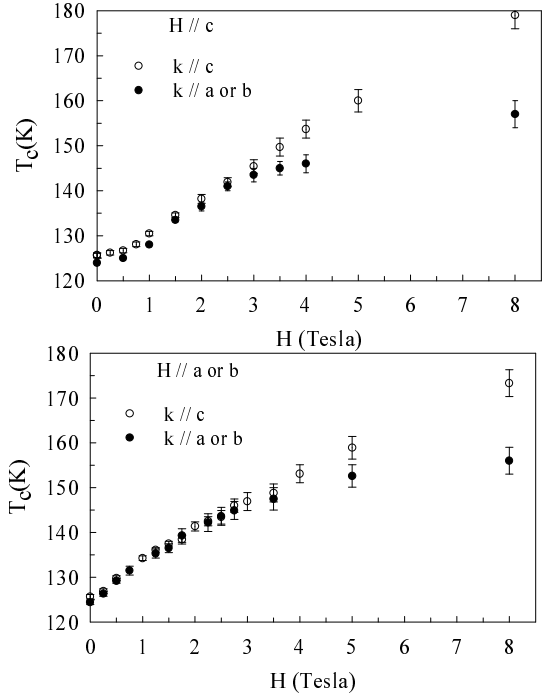}} \caption{Magnetic phase diagram
for both magnetic field orientations: T$_c$(H) is determined from
the maximum slope of the velocity variation in the critical
region for both wave propagation configurations.} \label{}
\end{figure}

Finally, the shift of the transition temperature T$_c$ as a
function of the magnetic field has also been investigated for
both wave propagation configurations. The deduced magnetic phase
diagram is shown in figure 8. The critical temperature T$_c$(H)
has been determined with the criterion used in zero magnetic
field. As expected, there is essentially no difference between
the T$_c$'s determined from in-plane and out-of-plane propagation
for both orientations of the magnetic field, but only for H $<$
3-4 Tesla. For higher field values, the T$_c$'s differ
surprisingly. We will thus discuss these field ranges separately.
When 0 $\leq$ H $\leq$ 3 Tesla, the magnetic phase diagram is not
dependent on the choice of the wave propagation but it clearly
depends on the field orientation. When the field is oriented
along one of the easy axes $a$ or $b$ (lower panel), T$_c$(H)
increases quasi-linearly with field up to 1.5 - 2 Tesla with a
rate of $\sim$ 8 K/Tesla; for higher fields, the rate gets smaller
around 6.3 K/Tesla. For the field oriented along the hard axis
$c$ (upper panel), the relation between T$_c$ and H is rather
quadratic at low field values as indicated by the positive
curvature noticed in Fig.8. At higher fields (H $>$ 1.5 Tesla), we
get a linear increase with a rate similar to the easy plane
orientation one. This slower increase of T$_c$ along the $c$ axis
is coherent with an anisotropy field around 1.5 Tesla as
determined previously from the field dependence of the low
temperature stiffening anomaly (Fig.5). For 3 $\leq$ H $\leq$ 8
Tesla, if T$_c$ is determined from the $c$ axis propagation, the
linear dependence on field observed below 3 Tesla is maintained
up to the maximum value of 8 Tesla. T$_c$(H) determined from
in-plane wave propagation is however smaller. Since the critical
region is much less affected by the fluctuations and domain walls
for wave propagation along the hard axis, the true T$_c$(H) must
be determined from this configuration; in the case of wave
propagation along one of the easy axis, fluctuations widen the
transition in such a way that T$_c$ appears to be underestimated.
The open circles of figure 8 are thus fully representative of the
true phase diagram for this compound.

\section{Conclusion}
\label{sec:3} We have identified in the layered
La$_{1.2}$Sr$_{1.8}$Mn$_2$O$_7$ CMR compound an important
coupling between longitudinal acoustic phonons and ferromagnetic
moments. In the ferromagnetic phase at low temperatures, this
coupling produces a stiffening anomaly on the ultrasonic velocity
from which the temperature profile of the order parameter, the
spin-spin correlation function, could be determined if domain
walls effects are taken into account. The field dependence of
this stiffening anomaly at low temperatures confirms the known
magnetic structure of this compound: an easy plane of
magnetization along $ab$ and a hard axis along $c$ with an
anisotropy field just above 1 Tesla. In the critical region, the
exponent of the temperature power law indicates that a 2D Ising
model is appropriate to describe the magnetization process. A
small anomaly was also observed for T$<$ 20 K but its origin
could not be identified. We have also observed above T$_c$ an
important softening of the velocity which confirms that 2D
ferromagnetic fluctuations are present in the paramagnetic state
over a wide temperature range. However, with our ultrasonic
measurements, we could not identify any effects due to the
presence of an antiferromagnetic order or fluctuations.

\begin{acknowledgement}
The authors acknowledge the technical assistance of M.
Castonguay. This work was supported by grants from the Fonds pour
la Formation de Chercheurs et l'Aide \`a la Recherche of the
Government of Qu\'ebec (FCAR) and from the Natural Science and
Engineering Research Council of Canada (NSERC).

\end{acknowledgement}

%
% BibTeX users please use
% \bibliographystyle{}
% \bibliography{}

\end{document}